\documentstyle[epsf]{l-aa}

\begin{document}

  \thesaurus{12.          % A&A Section 12: Physical processes
              (08.14.1;   % Stars: neutron
               02.04.1;   % Dense matter
               02.13.1)   % Magnetic fields
            }

% *************************************************************************
%                                 TITLE
% *************************************************************************
\title{Direct Urca process in strong magnetic fields \\
       and neutron star cooling}

\author{D.A.\ Baiko and D.G.\ Yakovlev
}
\institute{
         A.F.\ Ioffe Physical Technical Institute,
        194021, Politekhnicheskaya 26, St.Petersburg, Russia}
\date{}
\maketitle
\markboth{D.A.\ Baiko and D.G.\ Yakovlev: Direct Urca process
          in strong magnetic fields}{}

% *************************************************************************
%                             ABBREVIATIONS
% *************************************************************************
\def\la{\;\raise0.3ex\hbox{$<$\kern-0.75em\raise-1.1ex\hbox{$\sim$}}\;}
\def\ga{\;\raise0.3ex\hbox{$>$\kern-0.75em\raise-1.1ex\hbox{$\sim$}}\;}

% *************************************************************************
%                               TEXT BODY
% *************************************************************************

\begin{abstract}
The effect of the magnetic field on the energy
loss rate in the direct Urca reactions is studied.
The general expression for
the neutrino emissivity at arbitrary
magnetic field $B$ is derived.
The main emphasis is laid on a case, in which the field
is not superstrong, and
charged reacting particles ($e$ and $p$) populate many Landau levels.
The magnetic field keeps the process operative if
$\Delta k / k_{Fn} \la N_{Fp}^{-2/3}$
($N_{Fp}$ is the number of the Landau levels populated by protons
and $\Delta k \equiv k_{Fn}-k_{Fp}-k_{Fe}$),
that is beyond the well--known
switch--on limit in the absence of the field, $\Delta k < 0$.
Cooling of
magnetized neutron stars with strong
neutron superfluid in the outer cores and nonsuperfluid
inner cores is simulated. The magnetic field near the
stellar center speeds up the cooling
if the stellar mass $M$ is slightly less than
the minimum mass $M_c$,
at which the direct Urca reaction becomes allowed for $B=0$.
If $B = 3 \cdot 10^{16}$~G, the affected mass range is
$M_c-M \la 0.1 \, M_c$,
while for $B = 3 \cdot 10^{15}$~G the range is
$M_c-M \la 0.015 \, M_c$.
This may influence a theoretical interpretation of
the observed thermal radiation as illustrated
for the Geminga pulsar.
The case of
superstrong magnetic fields ($B \ga 10^{18}$~G),
such that $e$ and $p$ populate
only the lowest Landau levels is briefly outlined.
\end{abstract}

% Section 1 ****************************************************
\section{Introduction}
The presence or absence of the direct Urca process
($ n \to e + p + \bar{\nu}_e$, $e + p \to n + \nu_e$)
in the core of a neutron star is the most
important issue of the stellar cooling.
If operative, it dominates the cooling at the neutrino
stage (age $t \la 10^5$ -- $10^6$ yr) being several orders
of magnitude more efficient than any other
neutrino emission process (e.g., Pethick, 1992).
However, the direct Urca can occur
under stringent conditions: one requires quite
high fraction of protons [$k_{Fn} \le k_{Fp}+k_{Fe}$, where
$k_{F\alpha} = (3 \pi^2 n_\alpha)^{1/3}$ is a Fermi momentum,
and $n_\alpha$ is a number density of particles of species $\alpha$]
in order to conserve momentum
in the reaction. Nevertheless, some equations of state (EOSs)
allow for that
(Lattimer et al., 1991). On the other hand, most realistic
EOS of dense matter (Wiringa et al., 1988)
predicts too small fraction of protons,
which decreases with growing density; consequently,
the direct Urca is forbidden in the entire neutron star core.

In this paper we study the possibility for the direct Urca
to be open in the presence of a magnetic field $B$,
if the proton fraction is too low to open the process at $B=0$.
The beta--decay and related reactions in strong
magnetic fields have been studied since late 1960's
(e.g., Canuto \& Chiu 1971, Dorofeev et al. 1985,
Lai \& Shapiro 1991, and references therein).
However, these results have been obtained under
various simplified assumptions (constant matrix elements,
non--degenerate nucleons, etc.) and do not give the
emissivity of the direct Urca reaction
in the neutron star cores.
Several works on the subject have appeared most recently.
Leinson \& P\'erez (1997) considered the
case of superstrong fields ($B \ga 10^{18}$ G),
in which electrons and protons
occupy only the lowest Landau levels.
They found that such fields relaxed
the requirement of high proton fraction,
and the direct Urca was always permitted leading to a rapid
cooling of a neutron star.
The case of superstrong fields was examined also by
Bandyopadhyay et al.\ (1998).
These authors found, by contrast, that the condition
``$k_{Fn} \le k_{Fp} + k_{Fe}$'' (quotation implies
that one should be careful with definition of the Fermi
momentum in superstrong fields)
still determined the direct Urca threshold,
and that the fields enhanced the neutrino
emissivity by 1 -- 2 orders of magnitude
in the permitted regime compared
to the standard value (Lattimer et al., 1991)
\begin{equation}
    Q_\nu^0 = 4 \cdot 10^{27} (n_e/n_0)^{1/3} ~ T^6_9~~
              {m_n^\ast m_p^\ast \over m_n m_p} ~~
    {\rm erg~cm^{-3}~s^{-1}}~.
\label{QQ0}
\end{equation}
Here $T_9=T/10^9$ K,
$n_e$ is the electron density, $n_0 = 0.16$ fm$^{-3}$,
$m_n^\ast$ and $m_p^\ast$ are nucleon effective masses
in dense matter,
$m_n$ and $m_p$ are their bare masses.

The present paper is organized as follows.

In Sect.\ 2 we obtain a general expression for the
neutrino emissivity $Q_\nu$
of the direct Urca process.

In Sect.\ 3 we concentrate on a realistic case,
in which the magnetic field is not extremely high (although
still high: $B \leq 3 \cdot 10^{16}$), and charged particles
populate many Landau levels.
We show that the field keeps the direct Urca open slightly
outside the region where $k_{Fn} < k_{Fp} + k_{Fe}$
(the standard, $B=0$, definition of $k_F$ applies).

In Sect.\ 4 we illustrate this result by a series
of cooling simulations of magnetized neutron stars
with no superfluid in the inner cores
and neutron superfluid in the outer cores.

Finally, in Sect.\ 5 we treat briefly the case
of super strong magnetic fields. We show that
the results by Leinson \& P\'erez (1997)
are basically correct, although not very accurate in details,
while those reported by Bandyopadhyay et al.\ (1998) are
inaccurate.

% Section 2 ******************************************************
\section{Quantum formalism}
In this section we obtain a general formula for
the neutrino energy emission rate in the
direct Urca reactions valid at any
magnetic field $B \ll 10^{20}$ G
(at higher fields protons become relativistic).
The calculation is done in the Born approximation
within the standard quantum mechanical framework
with weak interactions described by the
Weinberg--Salam--Glashow theory.
We adopt the conventional assumption
that reacting electrons are relativistic, while
protons and neutrons are nonrelativistic.
All these particles are strongly degenerate.
The rate of transition from an initial
state $|i\rangle$ to a final state $|f\rangle$ is
$Q_{fi} = 2 \pi \hbar^{-1} V^{-1} |\langle f |{\cal H}| i\rangle|^2
\delta ({\cal E}_f - {\cal E}_i)$, where
$V$ is the normalization volume and
${\cal H}$ describes
weak interaction in the
second--quantization formalism. For the neutron decay, we have
($c=\hbar=1$)
\begin{equation}
      {\cal H} = {G \over \sqrt{2}} \, \int_{(V)} {\rm d} {\bf r} \,\,
      \hat{\psi}^\dagger_p
      (\delta_{\mu 0} - g_A \delta_{\mu i} \sigma_i) \hat{\psi}_n \,
      \hat{\bar{\psi_e}} \gamma^\mu (1+\gamma^5)
      \hat{\psi}_\nu.
\label{H2q}
\end{equation}
In this case, $G = G_F \cos{\theta_C}$,
$G_F = 1.436 \cdot 10^{-49}$ erg cm$^3$
is the Fermi weak coupling constant,
$\theta_C \approx 13^{\rm o}$ is the Cabibbo angle,
$g_A = 1.261$ is the axial--vector coupling constant,
$\sigma_i$ is a Pauli matrix, and $\gamma^\mu$ is
a Dirac matrix.
Finally, $\hat{\psi}_\alpha$
is a field operator in the coordinate representation, i.e.,
the expression of the form $\sum_{q} \hat{a}_q \psi_q({\bf r})$,
where $q$ denotes full set of quantum numbers, $\hat{a}_q$
is an annihilation operator, and $\psi_q({\bf r})$
is an eigenstate.

To evaluate the total neutrino energy loss rate
(emissivity) we need
to sum $Q_{fi}$ times the energy of the newly born
antineutrino over all initial and final states.
First of all,
we can sum the matrix element $|\langle f |{\cal H}| i\rangle|^2$
over the electron spin states. The energy--conserving delta--function
is not affected by this summation, since all, except the lowest,
electron states are spin--degenerate. Choosing the Landau
gauge of the vector potential
${\bf A} = (-By,0,0)$ and performing a tedious calculation, we get
\begin{eqnarray}
   \sum_{\sigma_e} |\langle f |{\cal H}| i\rangle|^2 &=&
   {(2 \pi)^2 M \over L_x L_z V^2} \, \delta (k_{nz}-k_{ez}-k_{pz}-k_{\nu z})
\nonumber \\
   &\times& \delta (k_{nx}-k_{ex}-k_{px}-k_{\nu x}),
\label{gen}
\end{eqnarray}
where
\begin{eqnarray}
     {M \over G^2} &=& \frac{1}{2} \,\, \delta_{s_p s_n} \, (1+g_A^2) \,
     \left[  \left(1-{k_{ez} \over \varepsilon_e} \right)
     \left(1-{k_{\nu z} \over \varepsilon_\nu} \right) \, F^{\prime 2}
     \right.
\nonumber \\
     &+&  \left.
     \left(1+{k_{ez} \over \varepsilon_e} \right)
     \left(1+{k_{\nu z} \over \varepsilon_\nu} \right) \, F^2  \right]
\nonumber \\
     &+& \delta_{s_p s_n} \, g_A \, s_p \,
     \left[ \left(1-{k_{ez} \over \varepsilon_e} \right)
     \left(1-{k_{\nu z} \over \varepsilon_\nu} \right) \, F^{\prime 2}
     \right.
\nonumber \\
     &-&  \left.
     \left(1+{k_{ez} \over \varepsilon_e} \right)
     \left(1+{k_{\nu z} \over \varepsilon_\nu} \right) \, F^2  \right]
\nonumber \\
     &+& 2 \,\, \delta_{s_p, 1} \, \delta_{s_n, -1} \, g_A^2 \,
     \left(1+{k_{ez} \over \varepsilon_e} \right)
     \left(1-{k_{\nu z} \over \varepsilon_\nu} \right) \, F^2
\nonumber \\
     &+& 2 \,\, \delta_{s_p, -1} \, \delta_{s_n, 1} \, g_A^2 \,
     \left(1-{k_{ez} \over \varepsilon_e} \right)
     \left(1+{k_{\nu z} \over \varepsilon_\nu} \right) \, F^{\prime 2}
\nonumber \\
     &+&
     \delta_{s_p s_n} \, (1-g_A^2) {k_{e \bot} \over
     \varepsilon_e} {{\bf k_\nu}{\bf q} \over \varepsilon_\nu q}
     \, F F' .
\label{M}
\end{eqnarray}
In these equations, $k_{\alpha i}$
is a cartesian component of a particle momentum,
$\varepsilon_\alpha$ is a particle energy, $s_p = \pm 1$ and $s_n = \pm 1$
are, respectively,
the doubled proton and neutron spin projections onto the
magnetic field direction, $k_{e \bot} = \sqrt{2bn}$,
$n$ is the electron Landau level number, $b \equiv |e|B$,
${\bf k}_\nu$ is the antineutrino wave vector, and
${\bf q} = (k_{nx}-k_{\nu x},k_{ny}-k_{\nu y},0)$.
$L_x$ and $L_z$ are the normalization lengths.
Finally, $F=F_{n',n}(u)$ and $F'=F_{n',n-1}(u)$ are the
Laguerre functions (e.g., Kaminker \& Yakovlev, 1981),
$n'$ is the proton Landau level number,
and $u=q^2 / (2b)$. If any index ($n$ or $n'$) is negative,
$F_{n,n'}(u)=0$.

The next step consists in integrating over the $x$ components
of proton and electron momenta, which specify
the $y$ coordinates of the Larmor guiding centers
of these particles. This operation gives
the factor $L_x^2 L_y b / (2 \pi)^2$ and removes
the second delta--function in Eq.\ (\ref{gen}).
Thus, we may write a general formula for the neutrino emissivity
(including the inverse reaction which doubles
the emission rate) as
\begin{eqnarray}
  Q_\nu &=& {2 b \over (2 \pi)^7} \sum_{nn's_p s_n} \int
          {\rm d}{\bf k}_n \, {\rm d}{\bf k}_\nu \,
          {\rm d}k_{pz} \, {\rm d}k_{ez}
\nonumber \\
        &\times&
          f_n \, (1-f_p) \, (1-f_e) \,
           \delta(\varepsilon_n-\varepsilon_e-
                  \varepsilon_p-\varepsilon_\nu) \, \varepsilon_\nu
\nonumber \\
        &\times&
           \delta (k_{nz}-k_{ez}-k_{pz}-k_{\nu z}) \, M.
\label{Qnu}
\end{eqnarray}
In this case, $f_\alpha =
(1+ \exp{[(\varepsilon_\alpha - \mu_\alpha)/T]})^{-1}$
is a Fermi--Dirac distribution,
and the particle energies are given by the familiar expressions:
\begin{eqnarray}
          \varepsilon_e &=& \sqrt{m^2_e + k_{ez}^2 + 2 b n},
\nonumber \\
          \varepsilon_p &=&  {k^2_{pz} \over 2 m_p^\ast} +
          \left[n' + \frac{1}{2}
          \left(1 - g_p s_p {m_p^\ast \over m_p} \right)
          \right] {b \over m_p^\ast},
\nonumber \\
          \varepsilon_n &=& {k^2_n \over 2 m_n^\ast} -
           {g_n s_n b \over 2 m_p}, \,\,\,\,\,\,\,
          \varepsilon_\nu = k_\nu,
\label{energs}
\end{eqnarray}
with the proton and neutron gyromagnetic
factors $g_p=2.79$ and $g_n=-1.91$.
%Kinetic energy of strongly interacting neutrons and protons
%in dense matter differs from that in vacuum,
%which is reflected by the appearance of the effective masses
%$m_n^\ast$ and $m_p^\ast$.
In principle, the factors $g_A$, $g_p$,
$g_n$ can be renormalized in dense matter which we
ignore, for simplicity.

Since the electron and proton distributions
are independent of signs of $k_{ez}$ and $k_{pz}$
we can simplify the expression for $M$ by
omitting the terms which would anyway yield zero
after the integration:
\begin{eqnarray}
     {M \over G^2} &=& \frac{1}{2} \,\, \delta_{s_p s_n} \, (1+g_A^2) \,
     \left[\left(1 + {k_{ez} \over \varepsilon_e}
     {k_{\nu z} \over \varepsilon_\nu} \right) \, F^{\prime 2}
     \right.
\nonumber \\
     &+& \left.
     \left(1+{k_{ez} \over \varepsilon_e}
     {k_{\nu z} \over \varepsilon_\nu} \right) \, F^2 \right]
\nonumber \\
     &+& \delta_{s_p s_n} \, g_A \, s_p \,
     \left[  \left(1+{k_{ez} \over \varepsilon_e}
     {k_{\nu z} \over \varepsilon_\nu} \right) \, F^{\prime 2}
     \right.
\nonumber\\
     &-& \left.
     \left(1+{k_{ez} \over \varepsilon_e}
     {k_{\nu z} \over \varepsilon_\nu} \right) \, F^2 \right]
\nonumber \\
     &+& 2 \,\, \delta_{s_p, 1} \, \delta_{s_n, -1} \, g_A^2 \,
     \left(1-{k_{ez} \over \varepsilon_e}
     {k_{\nu z} \over \varepsilon_\nu} \right) \, F^2
\nonumber \\
     &+& 2 \,\, \delta_{s_p, -1} \, \delta_{s_n, 1} \, g_A^2 \,
     \left(1-{k_{ez} \over \varepsilon_e}
     {k_{\nu z} \over \varepsilon_\nu} \right) \,  F^{\prime 2}
\nonumber\\
     &+&
     \delta_{s_p s_n} \, (1-g_A^2) \, {k_{e \bot} \over
     \varepsilon_e} \, {{\bf k_\nu}{\bf q} \over \varepsilon_\nu q} \, F F'.
\label{Ms1}
\end{eqnarray}
Keeping $k_{\nu z}$ in the $z$ component of the momentum conserving
delta--function in Eq.\ (\ref{Qnu}) would lead to a subtle thermal effect:
it would mollify the resulting functions on a temperature
scale. We will not pursue the accurate description
of the effect here, both because the calculation
would be quite complex, and because the temperature scale
is assumed to be small.
Therefore, we will neglect the neutrino momentum in the
delta--function and, for the same reason, omit it from
the definition of the vector {\bf q}. Then, using the isotropy
of the neutron distribution, we can further simplify
the expression for $M$:
\begin{eqnarray}
     {M \over G^2} &=& 2 g_A^2 \,
       \left( \delta_{s_p, 1} \, \delta_{s_n, -1} \, F^2
     + \delta_{s_p, -1} \, \delta_{s_n, 1} \, F^{\prime 2} \right)
\nonumber \\
     &+&  \frac{1}{2} \,\, \delta_{s_p s_n} \, (1+g_A^2) \,
     \left( F^{\prime 2} + F^2 \right)
\nonumber\\
     &+& \delta_{s_p s_n} \, g_A \, s_p \,
     \left( F^{\prime 2} -  F^2 \right),
\label{Ms2}
\end{eqnarray}
where the functions $F$ and $F'$ depend now on
$u=$ \\
$(k_{nx}^2+k_{ny}^2)/(2b) \equiv k_{n \bot}^2/(2b)$.

Notice that the results of this and subsequent
sections are equally valid for direct Urca processes
involving hyperons. The results for hyperons are
easily obtained by changing the values
of reaction constants ($g_A$, etc.) as described,
for instance, by Prakash et al.\ (1992).

% Section 3 *****************************************************
\section{Quasiclassical case}
{\it (a) General treatment and limit $B \to 0$.}
First of all consider the most realistic case
of not too high magnetic fields, in which electrons and
protons populate many Landau levels.
In this case the transverse wavelengths of
electrons and protons are much smaller than their
Larmor radii. Thus the situation may be referred to
as quasiclassical, and corresponding techniques apply.
If the main contribution
comes from large $n$ and $n'$, the difference between
$F^2$ and $F^{\prime 2}$ can be neglected.
Moreover, we can neglect the contributions
of magnetic momenta of particles to their energies.
Thus, we can pull all the functions of energy
out of the sum over $s_n$ and $s_p$, and evaluate the latter
sum explicitly:
\begin{equation}
    \sum_{s_n s_p} M = 2 \, G^2 \, (1+3 g_A^2) \, F^2.
\label{Msfin}
\end{equation}
Inserting this into Eq.\ (\ref{Qnu}),
and integrating over orientations of neutrino momentum
we get
\begin{eqnarray}
  Q_\nu &=& {16 \pi G^2 \, (1+3 g_A^2) \, b \over (2 \pi)^7} \int
          {\rm d} \varepsilon_\nu \,
          {\rm d}{\bf k}_n \, {\rm d}k_{pz} \, {\rm d}k_{ez}
          \, \varepsilon_\nu^3
\nonumber \\
         &\times& \sum_{nn'} \, F^2_{n',n} (u)\,
          f_n \, (1-f_p) \, (1-f_e)
\nonumber \\
         &\times&
          \delta(\varepsilon_n-\varepsilon_e-
                 \varepsilon_p-\varepsilon_\nu) \,
          \delta (k_{nz}-k_{ez}-k_{pz}) ,
\label{Qnu2}
\end{eqnarray}
where
\begin{equation}
       \varepsilon_p = {k_{pz}^2 + k_{p \bot}^2 \over 2 m_p^\ast},
       \,\,\,\,\,\,\,
       \varepsilon_n = {k_n^2 \over 2 m_n^\ast},
\label{en2}
\end{equation}
$k_{p \bot} = \sqrt{2bn'}$,
while the electron energy is still given by Eq.\ (\ref{energs}).

If the magnetic field is not too large
the transverse electron and proton momenta, $k_{e\bot}^2$ and
$k_{p\bot}^2$, are sampled
over a dense grid of values, corresponding to integer
indices $n$ and $n'$. Thus, the double
sum in Eq.\ (\ref{Qnu2}) tends
to the double integral over $k_{e \bot}^2$
and $k_{p \bot}^2$,
\begin{equation}
    2b \sum_{nn'} F^2_{n',n}(u) \ldots \to
    \int {\rm d}k_{e \bot}^2 {\rm d}k_{p \bot}^2 \,\,
    {\cal F} \ldots,
\label{sum-int}
\end{equation}
where ${\cal F}$ represents the small--$b$ asymptote
of the function
\begin{equation}
        {1 \over 2b} F^2_{Np,Ne} \left( u \right),
\label{funspec}
\end{equation}
with $N_p \equiv k_{p \bot}^2/(2b)$ and
$N_e \equiv k_{e \bot}^2/(2b)$.

After replacing the double sum by the double integral
Eq.\ (\ref{Qnu2}) can be considerably simplified.
Note, that
${\rm d}k_z \, {\rm d}k_{\bot}^2 = {\rm d}{\bf k}/ \pi =
2 m^\ast \, {\rm d}\varepsilon \, k \, \sin{\theta} \, {\rm d}\theta$,
where $\theta$ is a pitch--angle. Then
the energy integral $\int {\rm d}\varepsilon_n \, {\rm d}\varepsilon_p \,
{\rm d}\varepsilon_e \, {\rm d}\varepsilon_\nu$ is taken
by assuming, that the temperature
scale is small and provides the sharpest
variations of the integrand. If so,
we can set $k=k_F=(3 \pi^2 n)^{1/3}$,
$\varepsilon = \varepsilon_F$ etc.
in all the other functions. In principle,
this assumption constraints the validity of
the quasiclassical approach.
We will come back to this point at the end of this section.
Finally, we integrate over the azimuthal angle of the neutron momentum,
and over its polar angle to eliminate the momentum
conserving delta--function, and obtain:
\begin{eqnarray}
    Q_\nu &=& Q_\nu^0 \times R_B^{\rm qc};
\nonumber \\
         Q_\nu^0 &=&
         {457 \pi \,  G^2 \,(1+3 g_A^2) \over
         10080} \, m_n^\ast \, m_p^\ast \, \mu_e \, T^6,
\nonumber \\
        R_B^{\rm qc} &=&
        2 \int \int^{1}_{-1} {\rm d}\cos{\theta_p} \, {\rm d}\cos{\theta_e}
        \, {k_{Fp} k_{Fe} \over 4b} \,
        F^2_{Np,Ne}(u)
\nonumber \\
        &\times&  \Theta(k_{Fn}-
        |k_{Fp} \cos{\theta_p}+k_{Fe} \cos{\theta_e}|),
\label{Qnu3}
\end{eqnarray}
where $Q_\nu^0$ is the field--free emissivity (\ref{QQ0}),
and the factor $R_B^{\rm qc}$ describes the effect of
the magnetic field.
$k_{p,e \bot} = k_{Fp,e} \sin{\theta_{p,e}}$,
and $k_{n\bot}^2$ is now given
by $k_{Fn}^2 - (k_{Fp} \cos{\theta_p}+k_{Fe} \cos{\theta_e})^2$;
$\Theta(x)=1$ for $x > 0$, $\Theta(x)=0$ for $x<0$.

The asymptotic behaviour of (\ref{funspec}) depends
on the relation between the argument of the function $F$
and its indices. In the small--$b$ case one can distinguish
three domains of these parameters: (I) $k_{n\bot}<|k_{p\bot}-k_{e\bot}|$;
(II) $|k_{p\bot}-k_{e\bot}|<k_{n\bot}<(k_{p\bot}+k_{e\bot})$;
and (III) $(k_{p\bot}+k_{e\bot})<k_{n\bot}$. In domains
(I) and (III) the asymptotes in question decay exponentially
when $k_{n\bot}$ departs from the
domain boundaries (which are the turning points
of corresponding quasiclassical equation, e.g.,
Kaminker \& Yakovlev, 1981).
In both cases the exponents are inversely proportional to $b$,
and both asymptotes
tend to zero as $b\to 0$, although nonuniformly in the vicinities
of the turning points. In domain (II), the asymptote
of the expression (\ref{funspec}) oscillates according to
\begin{eqnarray}
         {\cal F} \approx
           {1 \over \pi pb} \cos^2 \Phi~, \,\,\,\,
           p = \sqrt{4 N_p N_e - (N_p + N_e - u)^2 }~,
\label{a-osc}
\end{eqnarray}
with a prefactor that is actually independent of $b$.
The cosine phase is
\begin{eqnarray}
       \Phi &=& - (1 + N_p) \,
       {\rm arg} (u + N_p - N_e, -ip)
\nonumber \\
       &+& N_e  \, {\rm arg}
       (u + N_e - N_p, ip)
       - {p - \alpha \over 2}~,
\label{osc-phas} \\
       {\rm e}^{i\alpha} &=& {p (N_p - N_e) +
                    i \left[(N_p - N_e)^2 -
                    u (N_p + N_e) \right]
                    \over 2 u \sqrt{N_p N_e}}~,
\nonumber
\end{eqnarray}
where arg$(z)$ is an argument of complex
number $z$ which lies in the range $[-\pi,\pi]$.
For integer indices, this formula coincides with Eq.\ (30)
in Kaminker \& Yakovlev (1981),
and it provides an accurate extension of ${\cal F}$
to non--integer indices.
One can easily show that a natural continuation of
$F^2_{Np,Ne}(u)/(2b)$ to this case is
given by the analytical
function $W^2_{km}(u)/[2b u \Gamma(N_e+1) \Gamma(N_p+1)]$,
where $W_{km}(u)$ is
the Whittaker function, $u=k^2_{n\bot}/(2b)$,
$k=(1+N_p+N_e)/2$, and $m=(N_e-N_p)/2$.
Equations (\ref{a-osc}) and (\ref{osc-phas})
are derived using accurate
small--$b$ asymptotes of $W_{km}(u)$, which, in turn,
could be derived from the integral representation of this
function by the saddle--point method.

This information is sufficient to check,
if $R_B^{\rm qc}$ reproduces the well--known step--function
$\Theta(k_{Fp}+k_{Fe}-k_{Fn})$
in the $b \to 0$ limit. If $k_{Fn}>k_{Fp}+k_{Fe}$,
we are always in domain (III) (since the $z$-momentum
is conserved), and $R_B^{\rm qc} \to 0$.
If, on the contrary, $k_{Fn}<k_{Fp}+k_{Fe}$ the integration in
(\ref{Qnu3}) covers all three domains, and the boundary
of domain (II) corresponds to vanishing
square root in Eq.\ (\ref{a-osc}).
The contribution from domains (I) and (III)
is again zero, while the integration over domain (II)
yields exactly $1$. To verify this, one can put
the rapidly oscillating $\cos^2{\Phi}$ equal to $1/2$
in Eq.\ (\ref{a-osc}), and
change the integration variables: $s=\cos{\theta_p}+\cos{\theta_e}$,
$t=\cos{\theta_p}-\cos{\theta_e}$. In all cases,
theta--function in Eq.\ (\ref{Qnu3}) plays no role.

{\it (b) Effect of the magnetic field in
the forbidden domain ($k_{Fn}>k_{Fp}+k_{Fe}$).}
In a finite magnetic field,
the border between the open and
closed direct Urca regimes
is expected to be smeared out over
some scale depending on the field strength.
This can be important for
neutron star cooling, since
the direct Urca process
can remain
the dominant energy loss mechanism
even if it
is suppressed exponentially by several orders of magnitude.

To investigate this possibility, one has to calculate
the emissivity $Q_\nu$ from Eq.\ (\ref{Qnu2}).
We have used two approaches,
one of which is
essentially quasiclassical and good
for a rapid computation, while the other is of quantum
nature --- more precise and time--consuming.

At the moment it is convenient to introduce two parameters,
$x$ and $y$, which characterize the reaction kinematics with respect
to the magnetic field strength:
\begin{equation}
   x  =  { k_{Fn}^2 - (k_{Fp}+k_{Fe})^2
           \over  k_{Fp}^2 \, N_{Fp}^{-2/3}}, \,\,\,\,\,\,
   y = {N_{Fp}}^{2/3},
\label{xy}
\end{equation}
where $N_{Fp}= k_{Fp}^2/(2b)$ is the
number of the Landau levels populated by protons.
The most interesting situation, from practical point of view,
occurs if $x$ is positive and not very large (say $x \la 10$), since
whenever $k_{Fn}$ significantly exceeds $k_{Fp}+k_{Fe}$,
the reaction is suppressed too strongly.
The first approach is based on Eq.\ (\ref{Qnu3}).
The main contribution to this integral
comes from the vicinity
of the line $\theta_p = \theta_e$, corresponding
to the least distance from $k_{n\bot}$ to domain (II).
Since this distance should not be large,
we may use in (\ref{Qnu3}) the small--$b$ asymptotic form of (\ref{funspec})
in the neighbourhood of the right turning point, i.e., for
$k_{n\bot} \approx k_{p\bot}+k_{e\bot}$. It is given,
for instance, by Kaminker \& Yakovlev (1981) and reads
\begin{eqnarray}
         F^2_{Np,Ne}(u) &\approx&
          {  (2b)^{2/3} (k_{p\bot} k_{e\bot})^{-1/3} \over
          (k_{p\bot} + k_{e\bot})^{2/3}  }
          \,\, {\rm Ai}^2(\xi)
\nonumber \\
          &=&
          { (\sin{\theta_p} \sin{\theta_e})^{-1/3} \over y
          \left( \sin{\theta_p} + \sin{\theta_e} \right)^{2/3} }
          \,\, {\rm Ai}^2(\xi),
\label{F-Ai}
\end{eqnarray}
\begin{eqnarray}
         \xi &=& { \left[k_{n\bot}^2 - (k_{p\bot}+k_{e\bot})^2 \right]
          (k_{p\bot} k_{e\bot})^{1/3} \over
          (2b)^{2/3} (k_{p\bot} + k_{e\bot})^{4/3}  }
\nonumber\\
         &\approx&
          \left[ x + 2y
          \left(1-\cos{(\theta_p-\theta_e)}\right) \right]
          {  \left( \sin{\theta_p} \sin{\theta_e} \right)^{1/3} \over
          \left( \sin{\theta_p} + \sin{\theta_e} \right)^{4/3} },
\nonumber
\end{eqnarray}
where Ai$(\xi)$ is the Airy function, defined as in
Abramowitz \& Stegun (1970),
and we have assumed $k_{Fp} = k_{Fe}$, which
implies absence of muons and hyperons in neutron--star matter.
At negative arguments, which correspond to domain (II),
the Airy function oscillates,
while for positive arguments, in domain (III),
it decreases exponentially approaching the asymptote
\begin{equation}
       {\rm Ai}^2 (\xi) \approx {1 \over 4 \pi \sqrt{\xi}} \,\,
       \exp{\left(-{4 \over 3} \xi^{3/2}\right)},~~~~~~~~~\xi \gg 1.
\label{Ai-as}
\end{equation}
Inserting the latter
equation into Eq.\ (\ref{Qnu3}),
we find, that at relatively large $x$
\begin{equation}
     R_B^{\rm qc} \approx \sqrt{y \over x + 12 \, y} \,\,\,
          {3 \over x^{3/2}} \,\,\, \exp{\left(-{x^{3/2} \over 3}\right)}.
\label{as-x!}
\end{equation}
At very large $x$, however, this asymptote should not be taken
too literally because Eq.\ (\ref{F-Ai}) ceases to be valid
far from the turning point.

On the other hand, under quasiclassical assumptions, we are
always interested in the case of $y \gg 1$.
If so, we can further
simplify Eq.\ (\ref{Qnu3}) at any $x$ using (\ref{F-Ai}),
and noting that all the functions can be expanded
around the line $\theta_p = \theta_e$. Then we obtain
\begin{eqnarray}
     R_B^{\rm qc} &=&  2^{-2/3}
           \int^{\infty}_{-\infty} {\rm d}s \int^{\pi}_0 {\rm d}\theta \,
          \sin^{2/3}{\theta} \,\,
          {\rm Ai}^2(\xi),
\label{Qnu4} \\
          \xi &=& {x + s^2 \over
          2^{4/3} \sin^{2/3} \theta}.
\nonumber
\end{eqnarray}
At $x=0$ this integral is taken analytically
and gives $R_B^{\rm qc}=1/3$.
By inserting Eq.\ (\ref{Ai-as}), one easily verifies,
that Eq.\ (\ref{Qnu4}) reproduces also the asymptote
(\ref{as-x!}) for $x \ll 12y$.

%
%******************************************************************
%                                                       FIGURE RBforb
\begin{figure}[ht]
\vspace{-0.5cm}
\begin{center}
\leavevmode
\epsfysize=8.8cm
%\end{center}
{\epsfbox{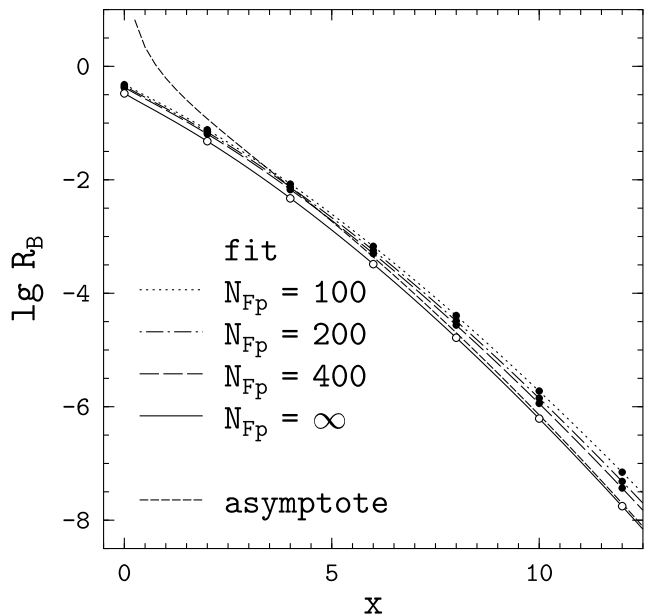}}
\end{center}
\vspace{-1cm}
\caption[ ]{
Logarithm of $R_B$ as a function of $x$
in the forbidden domain $\Delta k \equiv k_{Fn}-k_{Fp}-k_{Fe} > 0$
for various values of
$N_{Fp}$. Open circles correspond to the
quasiclassical approach ($R_B^{\rm qc}$), insensitive to the value of
$N_{Fp}$, while solid circles
show the quantum numerical results.
The short--dash line is the asymptote (\ref{as-x!}). Other
curves are calculated from the fitting formula (\ref{fitforb}).
}
\label{RBforb}
\end{figure}
%
%******************************************************************
%

Finally, we have calculated the factor $R_B^{\rm qc}$, using both
prescriptions (\ref{Qnu3}) + (\ref{F-Ai}), and
(\ref{Qnu4}) at $N_{Fp} \equiv y^{3/2} = 100, \, 1000, \, 10000$
and $x$ from 0 to 20.
In both these cases we have obtained
identical results. This indicates that for such combinations
of $x$ and $y$, the $y$--dependence of $R_B^{\rm qc}$ is insignificant.
These results are plotted in Fig.\ 1 by open circles.
The short--dashed curve represents the asymptote (\ref{as-x!}).

In the quasiclassical approach
we have made two approximations: firstly, we have replaced
the sum by the integral at finite $b$, and, secondly,
we substituted the asymptote in the form of the Airy function
for the function $F$. To assess the quality of both assumptions
we have used a quantum approach based directly on
Eq.\ (\ref{Qnu2}).

In this approach,
if one transforms the integration over $k_z$ to that
over $\varepsilon$ in a straightforward manner,
the integrand becomes singular
(the denominators of the form $\sqrt{\varepsilon - \varepsilon_n}$
appear). If $T \to 0$ the quantity $Q_\nu/T^6$
diverges at integer $N_{Fp}$. For nonzero $T$
this quantity remains convergent but oscillates as a function
of $B$ and/or density. These oscillations are quite
familiar and appear in many studies (magnetization, electrical
and thermal conductivities, etc., Landau and Lifshitz, 1986).
They are associated with population
of the Landau levels by electrons and
protons due to variation of plasma parameters.

If $T$ is larger than the energy spacing between the Landau levels
the oscillations are washed out, and a smooth curve emerges.
This regime requires very accurate integration over
particle momenta in order to include thermal effect.
We were able to perform it only for rather small $N_{Fp} \la 20$
and do not report these results here.
In a more important case of lower temperatures
the summation over Landau levels and energy integration
are independent. The actual
neutrino emissivity does oscillate but the quantity
of practical significance is the emissivity
averaged over the oscillations (a smooth curve again).
We call this approach quantum since it involves
the summation over Landau levels explicitly.
In this way we have calculated
the emissivity  and smoothed the oscillations
artificially by two different methods.
We have checked that both methods yield nearly
identical smoothed $Q_\nu$.
Simple consideration (see the end of this section)
shows that this (nonthermal quantum) approach is
valid for $T< {^3{\sqrt{N_{Fp}}}} \omega_B$,
or, equivalently,
$N_{Fp}^{2/3} < \mu_p/T$, where $\omega_B$ is the proton
gyrofrequency and $\mu_p$ is the proton chemical potential.

The results of these calculations are presented
in Fig.\ 1 for $N_{Fp} = 100, \, 200$, and 400
by solid circles. It is seen that with increasing $N_{Fp}$
the quantum factor $R_B$ tends to the quasiclassical
one $R_B^{\rm qc}$.

We have also found the fit
expression that describes accurately (Fig.\ 1) the quantum
calculations for $0 \leq x \leq 20$ and $N_{Fp} \geq 100$
and reproduces the quasiclassical curve
in the limit $N_{Fp} \to \infty$:
\begin{eqnarray}
       R_B &=& {(3 x + 6.800) \, R
             \over (x_c + 6.800) (3 + x \sqrt{12})} \,
             \exp{\left(-{x_c \over 3}\right)},
\label{fitforb} \\
       R~ &=& P_1 + P_2 {\rm e}^{P_3 x} + P_4 x,
\nonumber \\
       P_1 &=& 1 + {3.801 \over N_{Fp}^{0.5280}}, \,\,\,\,\,
       P_2 = 1 + {2.242 \over N_{Fp}^{0.3498}} - P_1,
\nonumber \\
       P_3 &=& 0.07221 + {1.751  \over N_{Fp}^{0.5001}}, \,\,\,\,\,
       P_4 = {1.551 \over N_{Fp}^{0.8128}},
\nonumber \\
       x_c &=& x \, \sqrt{x + 0.4176} - 0.04035 \, x.
\nonumber
\end{eqnarray}

{\it (c) Effect of the magnetic field in
the permitted domain ($k_{Fn}<k_{Fp}+k_{Fe}$).}
Since $R_B<1$ at $x=0$,
one may expect that the magnetic field has a non--trivial
effect on the neutrino emissivity in the permitted
domain. This appears to be true. Applying
the quasiclassical approach in the form of Eq.\ (\ref{Qnu4})
at negative $x$, we obtain for $R_B^{\rm qc}$ an oscillating curve,
shown in Fig.\ 2 by the open circles.
Using the more accurate
quantum approach we get a series of the curves for
$N_{Fp} = 100, \, 200$, and 400, the curve with highest
$N_{Fp}$ being again the closest to the quasiclassical result.
These oscillations
are of quasiclassical nature and have nothing in common
with the quantum oscillations discussed above.
From the practical point of view, they
are not very important, as they hardly have
any noticeable effect on the neutron star cooling.
Note, that the results for $-20 \leq x \leq 0$
and $N_{Fp} = \infty$ are accurately fitted by the expression
(solid curve in Fig.\ 2)
%
%     x:=abs(x);
%     FitFun:=1-1/Pow(par[1]+x,par[5])
%     *cos((x*par[8]+par[2]*x*Pow(x,par[6])+par[4])/(1+par[3]*Pow(x,par[7])));
%par[1] :=  5.815780E-0001;
%par[2] :=  8.452738E-0001;
%par[3] :=  1.437753E+0000;
%par[4] :=  1.211061E+0000;
%par[5] :=  1.191911E+0000;
%par[6] :=  1.532934E+0000;
%par[7] :=  1.209077E+0000;
%par[8] :=  4.822502E-0001; x=-0.05, -0.125 added
% Ndat =  83     rms error = 0.0101
%max error = 0.0262  at y= -2.500000E-0001
%
\begin{eqnarray}
       R_B &=& 1 - { \cos \varphi \over (0.5816 + |x|)^{1.192}},
\label{fitperm} \\
       \varphi &=& { 1.211 + 0.4823 |x| + 0.8453 |x|^{2.533}
                    \over 1 + 1.438 |x|^{1.209}}.
\nonumber
\end{eqnarray}
%

%
%******************************************************************
%                                                       FIGURE RBperm
\begin{figure}[ht]
\vspace{-0.5cm}
\begin{center}
\leavevmode
\epsfysize=8.8cm
%\end{center}
{\epsfbox{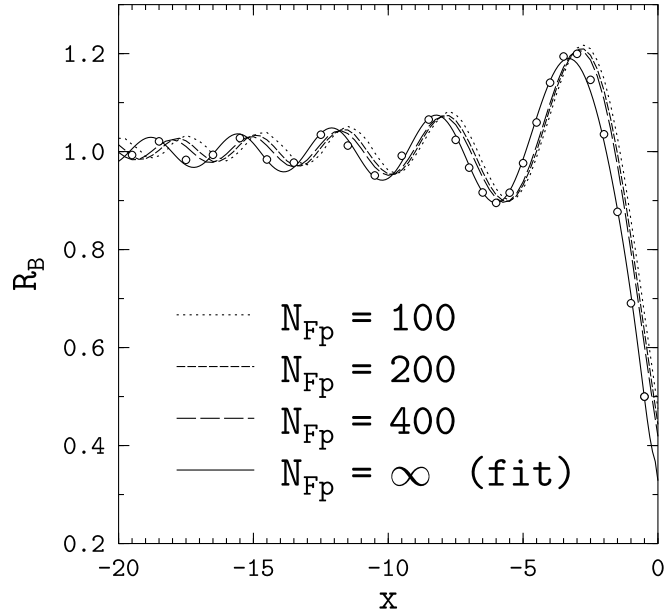}}
\end{center}
\vspace{-1cm}
\caption[ ]{
Factor $R_B$ in the permitted domain $\Delta k<0$
as a function of $x$ for various values of
$N_{Fp}$. Open circles correspond to the quasiclassical
approach (\ref{Qnu4}), solid line is calculated from the
fitting formula (\ref{fitperm}), and the other curves
represent the quantum numerical results.
}
\label{RBperm}
\end{figure}
%
%******************************************************************
%

To summarize, we remind that the direct Urca
reaction in the $B=0$ case is operative if the momentum
excess $\Delta k \equiv k_{Fn} - k_{Fp} - k_{Fe} < 0$.
In the presence of the magnetic field, the condition
becomes less stringent, and the reaction becomes quite efficient
as long as $x \la 10$, i.e., $\Delta k / k_{Fn}
\la N_{Fp}^{-2/3}$. If, for instance, $B = 10^{16}$ G,
and the density of matter is near the direct
Urca threshold, one typically has $N_{Fp} \sim 300$ and
$\Delta k / k_{Fp} \la 1/25$.

Notice that the field--free direct Urca process can also be allowed
beyond the domain $\Delta k < 0$ due to the
$thermal$ smearing of the Fermi surface. If $B=0$ and $\Delta k > 0$
the reaction rate can be written as $Q_\nu = Q_\nu^0 \times R_T$,
where $R_T$ may be expected to be
$\propto \exp (-v_{Fp} \, \Delta k / T)$. Thus the thermal effect
extends the reaction to the domain where
$\Delta k / k_{Fp} \la T/\mu_p$. The smearing is
clearly determined by the proton degeneracy parameter,
$\mu_p/T$, which is typically about 300 for $T \sim 10^9$~K.
The magnetic field effect is more important than the
thermal effect if $N_{Fp}^{2/3} \la \mu_p/T$
that can often be the case in the inner cores of neutron stars.

% Section 4. ***************************************************
\section{Cooling of magnetized neutron stars}
In this section we illustrate the
above results by cooling simulations.
Let us consider
a set of neutron star models with fixed equation
of state but varying central density and
magnetic field strength. Specifically,
we take the EOS
by Prakash et al.\ (1988) with the compression modulus
$K_0=180$ MeV and the symmetry energy in the form suggested
by Page \& Applegate (1992). It is assumed that
matter in the stellar core consist of neutrons, protons
and electrons (no muons and hyperons).
The EOS predicts
the monotonous increase of the proton fraction with increasing
mass density. Therefore,
if the central density $\rho_c$ is higher than the certain threshold
density $\rho_{\rm crit}$
(corresponding to proton fraction $x_p = n_p/n_b = 1/9$),
or, equivalently, the stellar mass $M$ is higher than
the certain threshold mass $M_c$
the direct Urca process becomes allowed in a central kernel
of the inner stellar core. For the
chosen EOS, the threshold parameters are
$\rho_{\rm crit} = 12.976 \cdot 10^{14}$ g cm$^{-3}$
and $M_c = 1.442 M_\odot$.

As pointed out by Page and Applegate (1992) the cooling history
of a neutron star in the field--free regime is
extremely sensitive to the stellar
mass: if the mass exceeds $M_c$
the star cools rapidly via the direct Urca process, while
the cooling of the low--mass star is mainly due to
the modified Urca process, and, therefore, is strongly delayed.
The effect of the magnetic field would be to speed up
the cooling of the star with a mass below
$M_c$, because the strong
field opens the direct Urca process
even if the $B=0$ condition $k_{Fn} \le k_{Fp}+k_{Fe}$
(or $x_p \ge 1/9$) is not reached.

The magnetic field in the neutron star core may evolve
on cooling time--scales. This may happen if
the electric currents supporting the field are located
in those regions of the core, where protons
as well as neutrons are nonsuperfluid. If so,
the electric currents transverse to the field
may suffer enhanced ohmic decay due to magnetization
of charged particles (e.g., Haensel et al., 1990).
The consequences of the decay would be
twofold. Firstly, if the strong field occupies a sufficiently
large volume of the core,
the decay produces an additional
heating,
which would delay the stellar cooling.
Secondly, the field decay would reduce
the direct Urca losses in the forbidden domain
decreasing the factor $R_B$.
If, however, the neutrons
are strongly superfluid, the enhanced decay is absent
(Haensel et al., 1990; {\O}stgaard \& Yakovlev, 1992),
and the ohmic decay time of the internal
magnetic field is typically larger than
the Universe age (Baym et al., 1969).

On the other hand, microscopic calculations
of superfluid
neutron gaps in the neutron star core suggest that the critical
temperatures are rather high at not too large densities
(say, below $10^{15}$ g cm$^{-3}$), but decrease
at higher densities (see, e.g., Takatsuka \& Tamagaki, 1997,
and references therein). Thus
the electric currents could persist in the outer stellar
core, where the neutron superfluid
is available, and the enhanced field--decay mechanism
does not work,
while the direct Urca reactions operate in the inner
core and are not subject to superfluid reduction.
This latter scenario we adopt. We assume the presence
of the magnetic field $B$ in the stellar kernel,
where the direct Urca process can be allowed, and
assume nonsuperfluid protons in the neutron star
core. Thus, the entire core is
nonsuperconducting,
and the magnetic field does not vary over the cooling time
(age $t \la 10^7$ yr) being frozen into the outer core.

Let us analyse the
cooling stage at which the neutron star interior is
isothermal ($t \ga 10^2$ -- $10^3$ yr).
The surface temperature $T_s$
(seen by a distant observer, i.e., the gravitational
redshift included) is then
determined by the heat transport through the neutron star
crust and is related uniquely to the internal
temperature.
We will use the cooling code described, for instance,
by Levenfish \& Yakovlev (1996), and Yakovlev et al.\ (1998).
The effects of General Relativity are included explicitly.
The neutrino luminosity is produced by
the standard neutrino reactions in the entire stellar core
complemented
by the direct Urca process in the inner core.
The effects of the neutron superfluidity
on the neutrino reactions and neutron heat capacity
in the outer core are taken into account as prescribed by
Levenfish \& Yakovlev (1996). In addition, we include
the neutrino emission due to triplet--state Cooper pairing of neutrons
(Yakovlev et al., 1998). The dependence of
the surface temperature on the internal stellar temperature
is taken from Potekhin et al.\ (1997) assuming
no envelope of light elements at the stellar surface.
To emphasize the effect of the internal magnetic fields on
the neutron star cooling we neglect the presence of
the surface magnetic fields and assume that the inner
field does not affect the relationship between the surface and internal
temperatures.

The density dependence of the neutron critical temperature
(triplet--pairing)
is given by the step--function:
$T_{cn}=10^{10}$ K
at $\rho<7.5 \cdot 10^{14}$ g cm$^{-3}$, and
$T_{cn}=0$
at $\rho \geq 7.5 \cdot 10^{14}$ g cm$^{-3}$
The resulting
cooling curves, $T_s(t)$, are insensitive to the initial inner
temperature
provided the latter is sufficiently high ($\ga 10^9$K).

The cooling curves are depicted in Figs.\ 3 and 4.
Dash line illustrates fast
cooling due to direct
Urca process allowed in a large
portion of the core at $B=0$.
While calculating this curve
we have used exact factors $R_B$ in all the
neutron--star layers where the field--free direct Urca is
either permitted or forbidden.
We have verified that the results are insensitive to
specific values of $R_B$ in the permitted domain.
In that domain, one can safely use the quasiclassical fit
(\ref{fitperm}) or even set $R_B=1$.
On the other hand,
even a huge internal field is unimportant
in the forbidden domain
(the curves for $B=0$ and $3 \cdot 10^{16}$ G
coincide): new regions
of the core, where the direct Urca is open by the field,
amount for a negligible fraction of the total neutrino luminosity.

%
%******************************************************************
%                                                       FIGURE 1292
\begin{figure}[ht]
\vspace{-0.5cm}
\begin{center}
\leavevmode
\epsfysize=8.8cm
%\end{center}
{\epsfbox{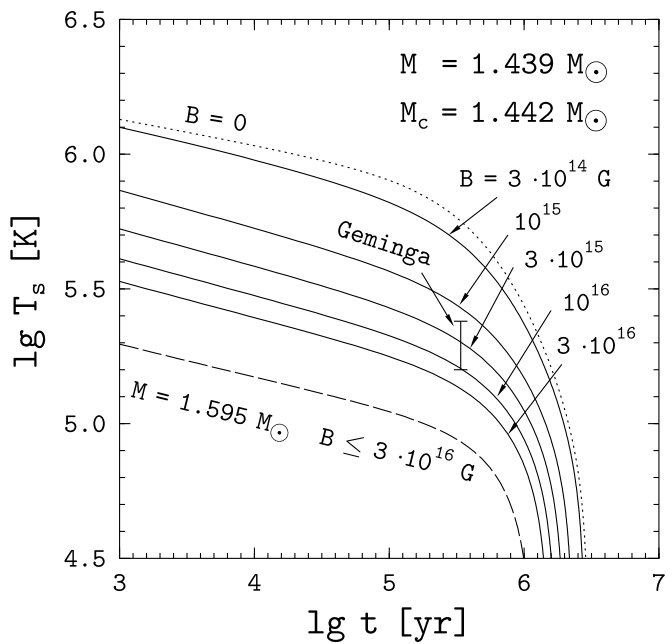}}
\end{center}
\vspace{-1cm}
\caption[ ]{Logarithm of the surface temperature as seen
by a distant observer as a function of neutron star age.
The dash curve is for a star of mass
$M=1.595 M_\odot$, well above the threshold mass $M_c$,
with magnetic field $0 \leq B \leq 3 \cdot 10^{16}$ G.
The dotted and solid curves are for the
$1.439 M_\odot$ star. Bar shows observations of
the Geminga pulsar (Meyer et al., 1994).
}
\label{1292}
\end{figure}
%
%******************************************************************
%

The upper dotted curves
are calculated for the stars
with masses 1.439 $M_\odot$ (Fig.\ 3) and 1.320 $M_\odot$
(Fig.\ 4) at $B=0$.
They represent the slow
cooling via the standard neutrino reactions
(the direct Urca is forbidden).
The solid curves illustrate the effect of the
magnetic field for the stars of the same masses.
If the stellar mass is slightly (by 0.2 \%) below
$M_c$ (Fig.\ 3) the cooling curve starts to deviate
from the standard one for not too high
fields, $B=3 \cdot 10^{14}$ G. Stronger fields,
$10^{15}$--$3 \cdot 10^{15}$ G produce the cooling
intermediate between the standard and rapid ones, while still
higher fields $B \ga 10^{16}$ G open
the direct Urca in a large fraction of the inner
stellar core
and initiate a nearly fully enhanced cooling. If, however, the mass
is by about 8\% below the threshold one (Fig.\ 4), only
a very strong field $B=3 \cdot 10^{16}$ G could keep
the direct Urca process
slightly open to speed up the cooling.

%
%******************************************************************
%                                                       FIGURE 110
\begin{figure}[ht]
\begin{center}
\leavevmode
\epsfysize=8.8cm
%\end{center}
{\epsfbox{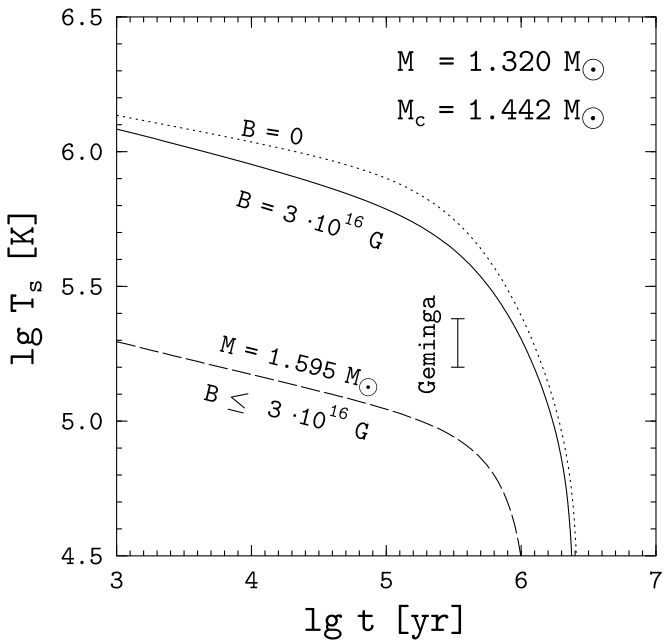}}
\end{center}
\vspace{-1cm}
\caption[ ]{Same as in Fig.\ 3. The dotted and solid curves
are for the $1.320 M_\odot$ star. The dash
line is the same as in Fig.\ 3.
}
\label{110}
\end{figure}
%
%******************************************************************
%

The results indicate that the magnetic field
in the very central stellar core can indeed
enhance the cooling provided the stellar mass
is close to the threshold mass $M_c$.
If $B = 3 \cdot 10^{16}$ G, the effect is significant
in a mass range
$M - M_c \la 0.1 \, M_c$.
For lower fields
the range becomes smaller. If, for instance,
$B = 3 \cdot 10^{15}$ G, the mass range becomes as narrow
as $M - M_c \la 0.015 \, M_c$.

%
%******************************************************************
%                                                       FIGURE mvsb
\begin{figure}[ht]
\vspace{-0.6cm}
\begin{center}
\leavevmode
\epsfysize=8.8cm
%\end{center}
{\epsfbox{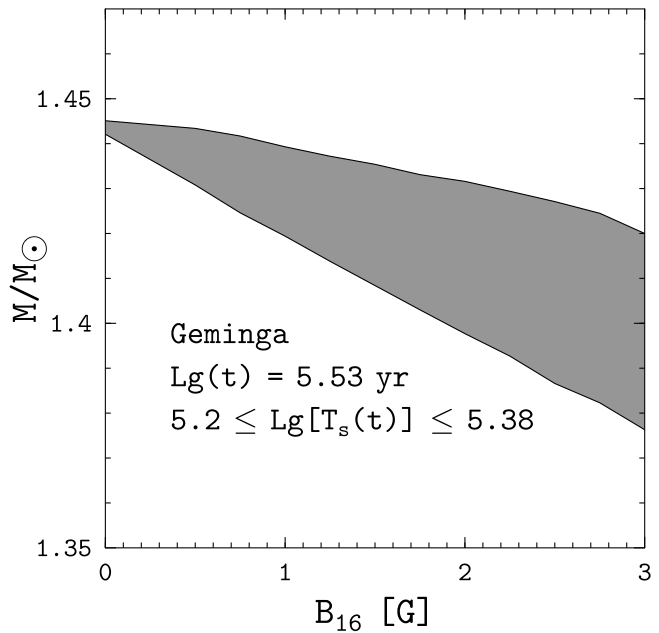}}
\end{center}
\vspace{-0.4cm}
\caption[ ]{The allowed mass range for the Geminga pulsar
as a function of the internal magnetic field
}
\label{mvsb}
\end{figure}
%
%******************************************************************
%

Our results can be used for interpretation of
observational data. By way of illustration,
consider observations of the thermal radiation
from the Geminga pulsar. Meyer et al.\ (1994)
fitted the observed spectrum by the set of
hydrogen atmosphere models. These
fits yield rather low non--redshifted effective surface
temperature $T_{\rm eff}=$ (2 -- 3)$\cdot 10^5$ K.
Introducing the appropriate
redshift factor $g=\sqrt{1 - R_g/R} \approx 0.8$
($R$ is the stellar radius and $R_g$ is the gravitational radius)
one gets
redshifted surface temperature $\lg T_s$ [K] $= 5.29 \pm 0.09$
($T_s=g \, T_{\rm eff}$). Adopting the dynamical Geminga's
age $t = 3.4 \times 10^5$ yr we can place
the Geminga's error bar in Figs.\ 3 and 4.
Let us use our cooling model (with possible strong
magnetic field $B$ near the stellar center unrelated to the
much weaker Geminga's surface magnetic field).
If $B=0$ Geminga is found
between the lines of standard ($M \leq M_c$) and
fast ($M > M_c$) cooling.
It is clear that tuning the mass slightly
above $M_c$ we can force the cooling curve to cross
the error bar. However, the mass range corresponding
to the bar width (Fig.\ 5) would be tiny (about $0.003 M_\odot$),
as the cooling rate is extremely sensitive
to $M$ in the domain just above $M_c$.
The narrowness of the confidence mass range
makes it fairly improbable that the Geminga's mass lies
in this range. Accordingly the suggested interpretation
of the Geminga's cooling is unlikely.
The situation becomes strikingly different in the presence
of the strong magnetic field.
The field shifts the confidence range of $M$ (faded area in Fig.\ 5)
below $M_c$ (cf.\ Fig.\ 3),
where variation of $T_s$ with $M$ at a given
$t$ is much smoother. This widens considerably
the confidence mass range.
At $B=10^{16}$ G it is about 0.02 $M_\odot$,
while at $B = 3 \cdot 10^{16}$ G it is about 0.04 $M_\odot$
so that the chances that the Geminga's mass
falls into this range become much higher.
This makes the proposed interpretation more plausible.

% Section 5. ***********************************************
\section{The case of superstrong magnetic fields}
Using the formalism
of Sect.\ 2 let us outline the main features of the
direct Urca reaction produced by electrons and protons occupying
the lowest Landau levels (with proton spins
aligned with the magnetic field). Notice, however,
that one needs superhigh
magnetic field, $B \ga 10^{18}$ G,
to force all electrons and protons into their ground
Landau levels. Let $n_e$ and $n_p$
be the number densities of these particles.
Contrary to the field--free Fermi momenta
valid at not too high fields and
used in Sects.\ 3 and 4,
the limiting
momenta along the superhigh magnetic field become field--dependent,
$k^0_{e,p} = 2 \pi^2 n_{e,p} / b$, $k^0_e \approx \mu_e$.
The distribution
of neutrons can be characterized by two Fermi momenta
for particles with spins along and against the magnetic field
$k_{Fn,s_n}$, $k_{Fn,1} < k_{Fn,-1}$. Our starting point
is Eq.\ (\ref{Ms2}), which reduces to
\begin{eqnarray}
      M &=& G^2 \, F^2 \,
          \left[{1 \over 2} (1-g_A)^2 \, \delta_{s_n,1} +
          2 g_A^2 \, \delta_{s_n,-1} \right].
\label{M00} \\
          F^2 &=& F^2_{00} = \exp \left(-{k^2_{n\bot} \over 2b} \right).
\nonumber
\end{eqnarray}
Let us insert it into Eq.\ (\ref{Qnu}), neglect $k_{\nu z}$
in the momentum--conserving delta function, and integrate
over the neutrino orientations and over
an azimuthal angle of the neutron momentum.
Then we convert the integrals over $k_{ez}$ and $k_{pz}$
into the integrals over particle energies,
take the standard energy integral, and perform
the integration over the neutron pitch--angle. We obtain
(the inverse reaction included)
\begin{eqnarray}
      Q_\nu &=& {457 \pi \, G_F^2 \, (1+3 g_A^2) \over 10080}
              m_n^\ast m_p^\ast \, \mu_e T^6
\nonumber \\
            &\times&
           {b \over k_p^0 k_e^0 \, (1+3 g_A^2)} \sum_{\alpha=\pm 1}
           \left[ \frac{1}{4} \, (1-g_A)^2 \,
           \Theta(u_{1 \alpha}) \, e^{-u_{1 \alpha}}
           \right.
\nonumber \\
           &+& \left. g_A^2 \,
           \Theta(u_{-1 \alpha}) \, e^{-u_{-1 \alpha}} \right],
\label{Q00}
\end{eqnarray}
where $2 b \, u_{s_n,\alpha} = k^2_{Fn,s_n} - (k^0_p + \alpha k^0_e)^2$,
and $\alpha = \pm 1$ corresponds to
two different reaction channels in which the electron and
proton momenta along the $z$-axis
are either parallel ($\alpha=1$) or antiparallel ($\alpha=-1$).
The step functions indicate that the channels are open
if $u_{s_n,\alpha} \geq 0$. The channel $\alpha=-1$
is always open in $npe$ dense matter with
the superstrong magnetic field, while
the channel $\alpha=1$ is open only if
$k_{Fn} \geq k^{0}_p + k^{0}_e$. The latter condition
is opposite to the familiar condition $k_{Fn} \leq k_{Fp}+k_{Fe}$
in the field--free case.
One has $\exp(-u_{s_n,\alpha}) \leq 1$ and
$Q_\nu \propto b^2$, but one cannot expect $Q_\nu$
to be essentially larger than the field--free emissivity
$Q^0_\nu$ as long as $B \la 10^{19}$ G.
Note also, that Eq.\ (\ref{Q00}) describes
the contribution of particles populating the
ground Landau levels to the emissivity
not only in the limit of superstrong fields.
For it to be valid at moderate fields one should substitute
number densities of $e$ and $p$ on the ground levels
for $n_e$ and $n_p$ in the definition of $k^0_{e,p}$.

Similar result for the superstrong magnetic fields
has been obtained recently
by Leinson and P\'erez (1997).
Nevertheless, their expression differs
from our in several respects.
Most importantly, the authors got
$(1+g_A^2)$ instead of $(1-g_A)^2$,
which substantially
overestimates the partial rate in the corresponding channel.
Much more different result under the same
assumptions was obtained by Bandyopadhyay et al.\ (1998).
In our notations,
the latter authors found the emissivity to be proportional
to $\Theta(-u_{11}) e^{-u_{11}}$, and
obtained the spurious
enhancement of the
neutrino emission and additional acceleration of the cooling
in a superstrong magnetic field. Their condition
which opens the reaction channel is opposite
to the actual one.

\section{Conclusions}
We have considered the neutrino emission produced by the
direct Urca process in the cores of neutron stars with
strong magnetic fields. We have derived the general
expression for the neutrino emissivity
[Eqs.\ (\ref{Qnu}) and (\ref{Ms1})], and
analysed it (Sect.\ 3) in the most important case, in which
the magnetic field is not too strong, and
the reacting electrons and protons populate many Landau
levels. We have shown that the magnetic field can
switch on the direct Urca process under the conditions,
in which the field--free process is strongly suppressed
by momentum conservation.
We have obtained the analytical fit expressions
(\ref{fitforb}) and (\ref{fitperm})
for this enhanced emissivity. We have performed
a series of neutron star cooling simulations for the model,
in which the star has a strong neutron superfluidity in
the outer core but no superfluidity in the inner core.
We have demonstrated,
that the magnetic fields $B \la 3 \cdot 10^{16}$ G
in the stellar center can strongly enhance the
neutrino luminosity and accelerate the cooling of
neutron stars with masses by
about 10 \% below
the threshold mass,
minimum mass of a star at which
the field--free direct Urca process
starts to be permitted.
Therefore, the cooling can be controlled by the
central stellar magnetic fields. The effect does not
require superstrong fields $B \ga 10^{18}$ G
analysed by different authors.

{\bf Acknowledgements:} D.A.B.\ is deeply thankful for
hospitality
at the Copernicus Astronomical center in Warsaw.
This work was supported by RFBR (grant 96-02-16870a),
RFBR-DFG (grant 96-02-00177G),
INTAS (grant 96-0542), and KBN (grant 2 P03D 014 13).

\end{document}